\documentclass[twocolumn,pra,psfig,showpacs,superscriptaddress]{revtex4}

\usepackage{graphicx}
\usepackage{times}
\usepackage{color}

\usepackage{graphics}

\usepackage{amssymb}

\begin{document}

\title{Magnetic effects on spontaneous symmetry breaking/restoration in a toroidal topology}

\author{L.M. Abreu{\footnote {luciano.abreu@ufba.br}}}
\affiliation{Instituto de F\'{\i}sica, Universidade Federal da
Bahia, 40210-340, Salvador, BA, Brazil}
\author{C.A. Linhares{\footnote {linharescesar@gmail.com}}}
\affiliation{Instituto de F\'{\i}sica, Universidade do Estado do Rio
de Janeiro, 20559-900, Rio de Janeiro, RJ, Brazil}
\author{A. P. C. Malbouisson{\footnote {adolfo@cbpf.br}}}
\affiliation{Centro Brasileiro de Pesquisas F\'{\i}sicas/MCTI,
22290-180, Rio de Janeiro, RJ, Brazil}
\author{J. M. C. Malbouisson{\footnote {jmalboui@ufba.br}}}
\affiliation{Instituto de F\'{\i}sica, Universidade Federal da
Bahia, 40210-340, Salvador, BA, Brazil}

\begin{abstract}
We study temperature and finite-size effects on the spontaneous symmetry breaking/restoration
for a scalar field model under the influence of an external magnetic field, at finite chemical
potential. We use the 2PI formalism and consider the large-$N$ limit. We find that there is a
minimal size of the system to sustain the broken phase,
which diminishes as the applied field increases but is independent of the chemical potential.
We analyze the critical curves and show that the magnetic field enhances the broken-phase
regions, while increasing the chemical potential leads to a diminishement of the critical
temperature.
\end{abstract}

\maketitle

\section{Introduction}

Field theories defined on spaces with some of its dimensions
compactified is interesting for several branches of theoretical
physics. They can be related, for instance, to studies of
finite-size scaling in phase transitions, to string theories or to
phenomena involving extra dimensions in high and low energy
physics~\cite{cardy,polchinski,panilinha,pani1,pani3,pani6,pani7,(g-2)NPB,claudio}. For a
Euclidean $D$-dimensional space, compactification of some
coordinates means that its topology is of the type $\Gamma
_{D}^{d}=(S^{1})^{d}\times \mathbb{R}^{D-d}$, with $1\leq d\leq D$,
$d$ being the number of compactified dimensions. Each of these
compactified dimensions has the topology of a circle $S^{1}$. We
refer to $\Gamma _{D}^{d}$ as a toroidal topology. Mathematical
foundations to deal with quantum field theories on toroidal
topologies are consolidated in recent
developments~\cite{AOP09,AOP11}. This provides a general framework
for  results from earlier works as for instance
in~\cite{Ademir,AMS,luc2,EPL12(1),PRD12,Emerson,Isaque}.

Here, in the framework of the two-particle irreducible (2PI)
formalism \cite{CJT,amelino} in the Hartree--Fock approximation, we
perform a study of magnetic effects for a field theory defined on a
toroidal topology. The main interest of field theories defined on
spaces with such a topology is that the simultaneous introduction of
temperature and finite-size effects is allowed in a natural way,
leading to size-dependent phase diagrams. We are particularly
interested in how a magnetic background affects the size-dependent
phase structure of the system; we present magnetic effects on
spontaneous symmetry restoration induced by both temperature and
spatial boundaries, at finite chemical potential. We will consider
the system with  a fixed squared mass parameter; within the toroidal
formalism, the model is valid for the whole domain of temperatures,
$0\leq T<\infty$.

\section{The 2PI formalism}

We consider the model described by the Lagrangian density
\begin{equation}
\mathcal{L} = \frac{1}{2}\partial_\mu\varphi^{*}
_{a}\partial^\mu\varphi_{a}+\frac{1}{2}m_0^{2}\varphi
^{*}_{a}\varphi _{a}+\frac{u}{4!}\,(\varphi ^{*}_{a}\varphi
_{a})^{2} \label{Lagrangiana}
\end{equation}
in a Euclidean $D$-dimensional spacetime, where $m_0$ and $u$ are
respectively the zero-temperature mass and the coupling constant in
the absence of boundaries, of external magnetic field and at zero
chemical potential. We consider the large-$N$ regime
where $N\rightarrow \infty$ and $u\rightarrow 0$ but
with $N u$ finite and fixed. To simplify the notation, we drop out
$a$-indices, summation over them being understood in field products.
We proceed to approach symmetry restoration for this model following
firstly the 2PI formalism~\cite{CJT,amelino} in the absence of
external field. In this case, the stationary condition for the
effective action, in the Hartree--Fock approximation, leads to the
gap equation
\begin{equation}
G^{-1}(\mathbf{x},\mathbf{x}^{\prime}) =
D^{-1}(\mathbf{x},\mathbf{x}^{\prime}) +
\frac{u}{2}\,G(\mathbf{x},\mathbf{x})\,\delta
^{D}(\mathbf{x}-\mathbf{x}^{\prime}), \label{GAP}
\end{equation}
with $\mathbf{x},\mathbf{x}^{\prime} \in \mathbb{R}^D$. The
Fourier-transformed propagators, $D({\bf k})$ and $G({\bf k})$, are
given by
\begin{equation}
D({\bf k}) = \frac{1}{{{\bf k}}^{2} + m_0^{2} + \frac{u}{2}
\phi^{2}}\,; \;\;\;\; G({\bf k}) = \frac{1}{{{\bf k}}^{2} + M^{2}}.
\label{D}
\end{equation}
Here, $\phi = \left\langle 0\left\vert
\sqrt{\varphi^{*}\varphi}\right\vert 0\right\rangle $ is the vacuum
expectation value of the quantum field $\varphi $ and $M$ is a
momentum-independent effective mass.

In the 2PI formalism, the gap equation corresponds to the stationary
condition and as such the effective mass depends on $\phi $ and
conveys all daisy and superdaisy graphs contributing to $G({\bf k})$
\cite{CJT,amelino}. Nevertheless, in order to investigate symmetry
restoration, we can take instead a particular constant value $M$ in
the spontaneously broken phase. Renormalization of the mass and of
the coupling constant can be performed with the
procedure described in Ref.~\cite{amelino}. Then, defining the
effective renormalized mass by ${\overline{m}}^{2}(\phi
)=-m_{R}^{2}+(u_{R}/2)\phi ^{2}$, where $m_{R}$ and $u_{R}$ are
respectively the renormalized mass and the renormalized coupling
constant, both at zero temperature and zero chemical potential, we
can write the gap equation, in momentum space, in the large-$N$
limit as~\cite{amelino,Isaque}
\begin{equation}
{\overline{m}}^{2}(\phi ) = M^{2} -
\frac{\lambda_{R}}{2}\frac{1}{(2\pi )^{D} }\int
d^{D}{k}\frac{1}{{{\bf k}}^{2}+M^{2}}, \label{lagrangiana1}
\end{equation}
where
\begin{equation}
\lambda_R=\lim_{N\rightarrow \infty,u_R\rightarrow 0}(Nu_R)
\label{lambdaR}.
\end{equation}
In the following, we will generalize this equation to include
effects of an external magnetic field as well as temperature,
chemical potential and size effects. We shall consider the constant
$\lambda_R$ as the physical renormalized coupling constant and focus
only on the correction of the mass.

\section{Two-point function in the presence of a  magnetic field }

In the presence of an external magnetic field, the Lagrangian
density in Eq.~(\ref{Lagrangiana}) becomes
\begin{equation}
\mathcal{L} =
\frac{1}{2}{{D}}^{\dagger}_\mu\varphi^{*}{{D}}^{\mu}\varphi
+\frac{1}{2}m_0^2\varphi^{*} \varphi
+\frac{u}{4!}\,(\varphi ^{*}\varphi )^{2},
\label{Lagrangiana1}
\end{equation}
where ${{D}} _\mu = \partial_{\mu} - ieA_{\mu}$ is the covariant
derivative and $A_{\mu}$ is the potential of the external gauge
field. We consider an uniform applied magnetic field
$H$ and choose a gauge such that $A=(0,x_1 H,0,0,\cdots)$. In this
case, the part of the Hamiltonian  quadratic in $ \varphi $ becomes,
after an integration by parts, $-\int d^{D}r\,\varphi ^{\ast}
\mathcal{D}\varphi $, where we have the differential
operator~\cite{lawrie1}
\begin{equation}
\mathcal{D}=\nabla ^2 -2i\omega x_1\partial _{x_2}-\omega
^{2}x_{1}^{2}-{m_{0}}^{2},
\end{equation}
with $\omega =eH$ being the cyclotron frequency. Thus the natural
basis to expand the field operators is the set of the normalized
eigenfunctions of the operator $\mathcal{D}$, the Landau basis.
Then, the free propagator can be written as~\cite{lawrie1}
\begin{equation}
{\mathcal{G}}(\mathbf{x},\mathbf{x}^{\prime }) = \int
\frac{d^{D-2}{q}\, d\kappa}{(2\pi )^{D-1}}\, \sum_{\ell =0}^{\infty
}\frac{\omega\, \xi_{\ell ,\kappa, \mathbf{q}}(\mathbf{x})\,
\xi_{\ell ,\kappa,\mathbf{q}}^{\ast }(\mathbf{x} ^{\prime
})}{\mathbf{q}^{2}+(2\ell +1)\omega +m_{0}^{2}}, \label{Propagator}
\end{equation}
with the Landau eigenfunctions given by
\begin{eqnarray}
\xi_{\ell ,\kappa, \mathbf{q}}(\mathbf{x}) & = &
\frac{1}{\sqrt{2^{\ell} {\ell}!}} \left( \frac{\omega}{\pi}
\right)^{\frac{1}{4}} e^{i
\mathbf{q}\cdot\mathbf{z}} \, e^{ i \omega \kappa x_2} \nonumber \\
& & \times\, e^{ - \frac{1}{2} \omega (x_1 - \kappa^2)} \,
H_{\ell}\left( \sqrt{\omega} (x_1 - \kappa) \right) ,
\end{eqnarray}
where $H_{\ell}$ denote the Hermite polynomials, with cartesian
coordinates, $\mathbf{x} = (x_1,x_2,\mathbf{z})$ and $ \mathbf{q}$
the $(D-2)$-dimensional momentum associated to the  vector
$\mathbf{z}$  in the gauge we choose.

Following Ref.~\cite{lawrie1}, we can extract the
non-translational-invariant phase of the propagator
(\ref{Propagator}) and write
\begin{equation}
{\mathcal{G}}(\mathbf{x},\mathbf{x}^{\prime}) = e^{i \omega (x_1 +
x_1^{\prime}) (x_2 - x_2^{\prime}) / 2}\, \bar{\mathcal{G}}
(\mathbf{x} - \mathbf{x}^{\prime}) ,
\label{G}
\end{equation}
where
\begin{eqnarray}
\bar{\mathcal{G}}(\mathbf{x} - \mathbf{x}^{\prime}) & = & \int
\frac{d^{D-2}{q}\, d\kappa}{(2\pi )^{D-1}}\, \sum_{\ell =0}^{\infty
}\frac{\omega\, e^{i \mathbf{q}\cdot (\mathbf{z} -
\mathbf{z}^{\prime})}}{\mathbf{q}^{2}+(2\ell +1)\omega +m_{0}^{2}}
\nonumber \\
& & \times\, \frac{1}{2^{\ell} \ell !} \sqrt{\frac{\omega}{\pi}}\,
e^{i \omega \kappa (x_2 - x_2^{\prime})}\,e^{-\frac{1}{4} [(x_1 -
x_1^{\prime})^2 + 4 \kappa]} \nonumber \\
& & \times\, H_{\ell} \left[ \sqrt{\omega} \left( \small{\frac12}
(x_1 - x_1^{\prime}) -\kappa \right) \right] \nonumber \\
& & \times\,H_{\ell}\left[ \sqrt{\omega}\left( - \small{\frac12}
(x_1 - x_1^{\prime}) -\kappa
\right) \right] \nonumber \\
& = & \int \frac{d^D k}{(2 \pi)^{D}}\, e^{i \mathbf{k}\cdot
(\mathbf{x} - \mathbf{x}^{\prime})}\,
\widetilde{\mathcal{G}}(\mathbf{k},\omega) \,.
\label{G1}
\end{eqnarray}
Taking the coincidence limit, $\mathbf{x}=\mathbf{x}^{\prime}$, and
using the orthonormality relations for the Hermite polynomials, we
find
\begin{equation}
\widetilde{\mathcal{G}}(\mathbf{k},\omega) = 2 \pi \delta(k_1)
\delta(k_2)
\sum_{\ell =0}^{\infty }\frac{\omega}{\mathbf{k}^{2}+(2\ell
+1)\omega +m_{0}^{2}} .
\label{G2}
\end{equation}
As a consequence, to introduce temperature and finite-size effects
in the gap equation we are restricted to perform
compactifications in the remaining $D-2$ coordinates; thus, to
consider both effects we have to consider a space-time with
dimension $D\geq 4$.

\section{Mass corrections in a toroidal space in the presence of an
external field}

To take into account finite-size and chemical-potential effects, we
consider first the changes introduced by the external field, given
in Eq.~(\ref {G2}) . To this end, let us remember that the parameter
$M$ is an effective mass taken as a constant. In such a case, the
changes due to the external applied constant magnetic field are
introduced via the minimal coupling, $\partial _{\mu }\rightarrow
D_{\mu }=\partial _{\mu }-ieA_{\mu }$ and we adopt the approximation
of neglecting the corrections arising from the vertices involving
the classical field $\phi$. This means that the integral $\int
d^{D}{k} \,({\bf k}^2+M^2)^{-1}$ in Eq.~(\ref{lagrangiana1}) should
incorporate the magnetic field as dictated by Eq.~(\ref{G2}), in
such a way that Eq.~(\ref{lagrangiana1}) takes the form
\begin{equation}
\overline{m}^{2}(\phi;\omega)=M^{2}-\frac{\omega
\lambda_{R}}{2}\sum_{\ell =0}^{\infty }\int \frac{d^{D-2}{q}}{(2\pi
)^{D-2}}\frac{1}{{\bf q}^{2} + M_{\ell }^{2}(\omega )},
\label{lagrangiana2}
\end{equation}
where $\overline{m}^{2}(\phi;\omega)$ is the $\omega$-dependent
effective renormalized mass and $M_{\ell }^{2}(\omega ) =
M^{2}+(2\ell +1)\omega$.

In the sequel we will obtain the generalization of
Eq.~(\ref{lagrangiana2}) in such a way as to include the toroidal
topology as well as the chemical potential. Restoration of the
symmetry will occur at the set of points in the toroidal space where
$ \overline{m}^{2}=0$.

We now proceed to generalize Eq.~(\ref{lagrangiana2}) to a theory
defined on a space with a toroidal topology. In the
$(D-2)$-dimensional system in thermal equilibrium  at temperature
$\beta ^{-1} $ and with compactification of $d-1$ spatial
coordinates (compactification lengths $L_{j}$, $j=2,3,\cdots d$). We
have $\mathbf{z}=({\tau},z_2,...,z_{d},\mathbf{w} )$, where $\tau $
corresponds to imaginary time and $\mathbf{w}$ is a
$(D-2-d)$-dimensional vector; the corresponding momentum is
$\mathbf{q}=(q_{\tau},q_2,...,q_{d},\mathbf{p})$, $\mathbf{p}$ being
a $(D-2-d)$-dimensional vector in momentum space.  We consider the
simpler situation of $d=2$, the system at temperature $\beta ^{-1}$
and one compactified spatial coordinate ($z_2\equiv z$) with a
compactification length $L_{2}\equiv L$. Then the Feynman rules
should be modified according to~\cite{AOP09,AOP11}
\begin{equation}
\int \frac{dq_{\tau }dq_{z}}{(2\pi)^{2}}f(q_{\tau},q_z,\mathbf{p})
\rightarrow \frac{1}{\beta L}\sum_{n_{\tau },n_z=-\infty }^{\infty
}f_{\omega_{n_{\tau}},\omega_{n_z}}(\mathbf{p}), \label{Matsubara}
\end{equation}
where the function $f_{\omega_{n_{\tau}},\omega_{n_z}}(\mathbf{p})$
is obtained from $f(q_{\tau},q_z,\mathbf{p})$ by the replacements
$q_{\tau }\rightarrow 2n_{\tau }\pi / \beta -i\mu$ and
$q_{z}\rightarrow 2n_{z}\pi /L_{}$, where $\mu $ is the chemical
potential. In this case, using Eq.~(\ref{Matsubara}), we can perform
a suitable generalization of the procedure in~\cite{amelino}, to
take into account finite-size, thermal and boundary effects in
Eq.~(\ref{lagrangiana2}). The integral over the $(D-2)$-dimensional
momentum in Eq.~(\ref{lagrangiana2}) becomes a double sum over
$n_{\tau }$ and $n_{z}$ together with a ($D-4$)-dimensional integral
over the remaining momentum $\mathbf{p}$.

Then, following steps similar to those
in~\cite{Ademir} and using dimensional regularization to perform the
integral, the renormalized $(T ,L,\mu ,\omega )$-dependent mass in
the large-$N$ limit can be written in the form
\begin{equation}
\overline{m}^{2}(T,L,\mu ,\omega )=M^{2}-\Sigma (T,L,\mu ,\omega ).
\label{msigma1}
\end{equation}
We introduce dimensionless parameters,
\begin{equation}
t=\frac{1}{M\beta},\,\chi=\frac{1}{ML}, \,\gamma=\frac{\mu}{M}, \,
\delta=\frac{\omega}{M^2}, \, \lambda=\frac{\lambda_R}{M^{4-D}}
\label{reduzidos}
\end{equation}
and the notation $M_{\ell}^{2}(\omega) \equiv M_{\ell}^{2}(\delta)
=M^{2} c_\ell ^2(\delta)$, where
\begin{equation}
c_{\ell}(\delta)=\sqrt{1+(2\ell+1)\delta}. \label{cl}
\end{equation}
Then  in terms of the dimensionless quantities we have,
\begin{eqnarray} \frac{\Sigma (t,\chi,\gamma
,\delta)}{M^2} = \lambda t \chi \delta\, \frac{\pi ^{(D-4)/2}}{8\pi
^{2}}\, \frac{\Gamma \left( s-\frac{D-4
}{2}\right) }{\Gamma \left( s\right) } \nonumber \\
\left. \times \sum_{\ell = 0}^{\infty} Z_{2}^{h^{2}_\ell} \left(
s-\frac{D-4}{2};t^2,\chi^2;b,0 \right) \right|_{s=1}, \label{sigma2}
\end{eqnarray}
where we define $b=i\gamma/2\pi t$ and $Z_{2}^{h^2}(\nu;\{a\},\{b\})
= \sum_{n,l=-\infty}^{\infty} \left[ a_1 (n - b_1)^2 + a_2 (l -
b_2)^2 + h^2 \right]^{-\nu} $ is an inhomogeneous Epstein--Hurwitz
zeta function~\cite{Elizalde}, with $h^{2}_\ell = (2\pi)^{-1}c_\ell
^2(\delta)$.

The Epstein--Hurwitz zeta functions have representations in the
whole complex $s$-plane in terms of modified Bessel functions of the
second kind $K_\nu$ \cite{Elizalde}; however, the first term of the
Epstein--Hurwitz function in this representation implies that the
first term in the correction to the mass is proportional to
$\Gamma\left( (4-D)/2\right)$, which is divergent for even
dimensions $D\geq 4$ \cite{Ademir}. This term is suppressed by a
minimal subtraction, leading to a finite effective renormalized
mass; notice that we call the quantities obtained after subtraction
of this polar term {\it renormalized} quantities, in the sense of
finite quantities, even if it is not a perturbative renormalization.
For the sake of uniformity, this polar term is also subtracted for
other dimensions, where no singularity exists, corresponding to a
finite renormalization. Notice also that the polar term which is
subtracted does not depend on $\beta $, $L$, $\mu $ and $\omega $.

This leads to the mass equation, written in terms of the above
dimensionless parameters,
\begin{eqnarray}
\frac{\overline{m}^2(t,\chi,\omega,\delta)}{M^2} = -1+{\lambda
\frac{1}{\pi (2\pi )^{
\frac{D-2}{2}}}}   \nonumber \\
\times \sum_{\ell =0}^{\infty }\left[ \sum_{n=1}^{\infty }\cosh
\left( \frac{\gamma n}{t}\right) \left(\frac{t}{n c_{\ell}(\delta)}
\right)^{\frac{D-4}{2}} K_{\frac{D-4}{2}}
\left( \frac{n}{t } c_{\ell}(\delta)\right) \right. \nonumber \\
\left. +\sum_{n=1}^{\infty }\left( \frac{\chi }{n c_{\ell}(\delta)}
\right) ^{\frac{D-4}{2}} K_{\frac{D-4}{2}}\left( \frac{n}{\chi }
c_{\ell}(\delta)\right) \right.  \nonumber \\
\left. + \,2\sum_{n,r=1}^{\infty }\cosh \left( \frac{\gamma n}{t
}\right) \left( \frac{1}{c_{\ell}(\delta) \sqrt{\frac{n^{2}}{
t^{2}}+\frac{r^{2}}{\chi ^{2}}}}\right)
^{\frac{D-4}{2}} \right. \nonumber \\
\left. \times K_{\frac{D-4}{2}}\left( c_{\ell}(\delta) \sqrt{
\frac{n^{2}}{t^{2}}+\frac{r^{2}}{\chi ^{2}}}\right) \right] .
\label{critica}
\end{eqnarray}

%%%%%%%%%%%%%%%
\begin{figure}[ht]
\includegraphics[{height=7.0cm,width=7.0cm}]{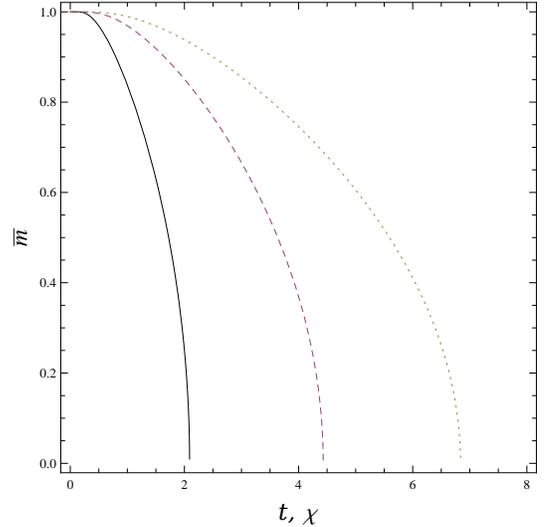}
\caption{Plot of corrected mass $\overline{m}$ in
Eq.~(\ref{critica}) in units of $M$ as a function of $t$ ($\chi$),
in the case of $ \gamma=0.0$, for the values of the reduced magnetic
field $\protect\delta =0.01$, $\protect\delta =0.5$ and $\delta=1.5$
(full,  dashed and dotted lines, respectively). We fix
$\lambda=1.0$, $ \chi \,(t) =0.001$.} \label{FigMag1}
\end{figure}
%%%%%%%%%%%%%%%
%%%%%%%%%%%%%%%
\begin{figure}[ht]
\includegraphics[{height=7.0cm,width=7.0cm}]{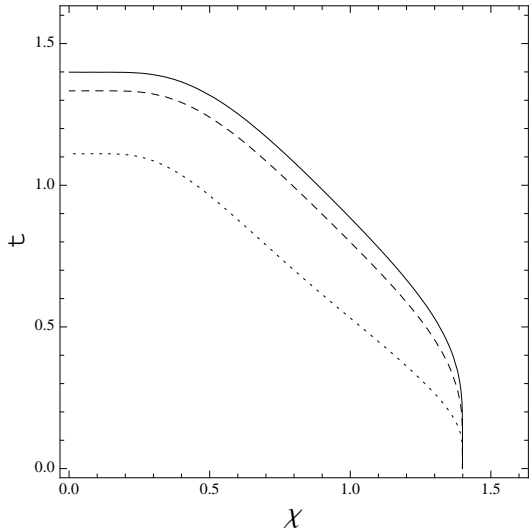}
\caption{Reduced critical temperature versus reduced inverse size
for a fixed value $\delta=1.0$ of the reduced applied field and
(right to left curves) for reduced chemical potentials $\gamma
=0.0,\,0.5,\,1.0$  (respectively full, dashed and dotted lines).  We
fix $\lambda=1.0$ The symmetry-breaking regions are below each
curve. } \label{FigMag2}
\end{figure}
%%%%%%%%%%%%%%%
%%%%%%%%%%%%%%%
\begin{figure}[ht]
\includegraphics[{height=7.0cm,width=7.0cm}]{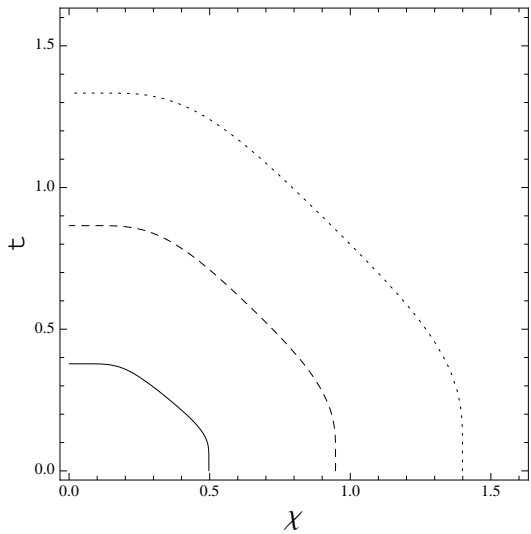}
\caption{Reduced critical temperature as a function of the reduced
inverse size of the system, for a fixed value of the reduced
chemical potential, $\gamma=0.5$ and the values of the reduced
magnetic field $\delta=0.01$, $\delta=0.5$ and $\protect\delta =1.5$
(respectively full, dashed and dot-dashed lines). We fix $\lambda
=1.0$. The symmetry-breaking regions are below each curve.}
\label{FigMag3}
\end{figure}
%%%%%%%%%%%%%%%
%%%%%%%%%%%%%%%
\begin{figure}[ht]
\includegraphics[{height=7.0cm,width=7.0cm}]{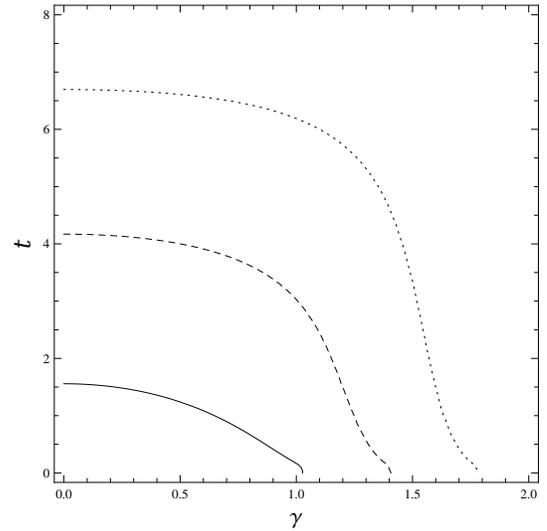}
\caption{Reduced critical temperature versus reduced chemical
potential for a fixed value $\chi=0.5$ and for $\delta
=0.01,\,0.5,\,1.5$  (respectively full, dashed and dotted lines). We
fix $\lambda =1.0$ The symmetry-breaking regions are below each
curve. } \label{FigMag4}
\end{figure}
%%%%%%%%%%%%%%%
\begin{figure}[ht]
\includegraphics[{height=7.0cm,width=7.0cm}]{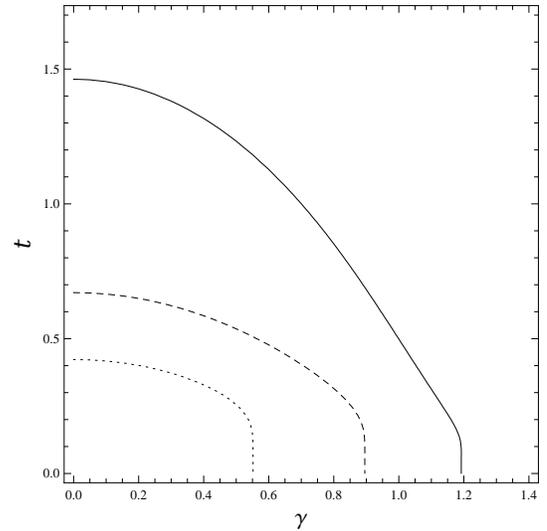}
\caption{Reduced critical temperature versus the reduced chemical
potential for three values of the reduced inverse size, $\chi=1.0$,
$\chi =2.0$ and $\chi =2.5$, for a fixed value of the reduced
applied field, $\delta =0.1$ (respectively full, dashed and
dot-dashed lines). We fix $\lambda=1.0$ The symmetry-breaking
regions are below each curve.  } \label{FigMag5}
\end{figure}
%%%%%%%%%%%%%%%

\section{Discussion}

Now we analyze the effects of the finite-size and external magnetic
field on the thermodynamic behavior of the system. In the general
situation, the resulting equation does not allow an algebraic
solution, and numerical evaluations are needed. For numerical
evaluations, we fix the value $\lambda=1.0$ and take several values
of the dimensionless parameters $t$, $ \chi $, $\delta$, and
$\gamma$.

In Fig.~\ref{FigMag1}, it is illustrated the behavior of the
effective corrected mass $\overline{m}$ defined in
Eq.~(\ref{critica}) for some values of the reduced magnetic field,
as a function of the reduced temperature $t$ or the reduced inverse
size $\chi$, at vanishing chemical potential. As it can be noted,
the behavior of $\overline{m}$ with respect to the quantities $t$
and $\chi$ are similar for $\gamma = 0$. The effective corrected
mass decreases as $t $, $\chi$ increase in the same way. We see that
for a given size, the critical temperature is higher for larger
values of the applied field. Conversely, the minimal size sustaining
the broken phase is smaller for higher values of the applied field.

To explore the results discussed above in more detail, we analyze
the critical behavior of the system. Criticality is attained for
$\overline{m}^{2}(t,\chi ,\mu ,\delta )=0$ in Eq.~(\ref{critica}).
The reduced critical temperature versus the reduced inverse size of
the system $\chi $ is plotted in Fig~\ref{FigMag2}. In this
situation we have fixed the value of the reduced applied field at
$\delta=1.0$, and the full, dashed and dotted lines represent the
critical lines for different values of the reduced chemical
potential: $\gamma =0.0,\,0.5,\,1.0$, respectively. We notice that
the critical temperature diminishes as the size of the system
increases, i.e. the broken phase is inhibited as the size of the
system decreases. Besides, this figure strongly suggests that there
is a minimal size of the system, $L_0$, (corresponding to a maximum
allowed  value of $\chi$, $\chi_0$), which is independent of the
chemical potential, below which the symmetry breaking disappears.
Also, the critical temperature depends on the density, in such a way
that, for fixed values of the thickness and of the applied field, it
is smaller for higher values of the chemical potential.

In Fig.~\ref{FigMag3}, the reduced critical temperature versus the
reduced inverse size of the system, $\chi $ is plotted for different
values of the reduced applied field: $\delta=0.01$, $\delta=0.5$ and
$\delta=1.5$, and for a fixed value of the reduced chemical
potential, $\gamma =0.5$ (respectively full, dashed and dot-dashed
lines). This figure shows that for higher applied fields the minimal
thickness of the system for which the transition exists is smaller.
In addition, we also see that the critical temperature is higher for
a higher applied field. It means that the broken phase for a thinner
system is favored as the magnetic field is increased;  the magnetic
field drives the system to the broken phase.

Another interesting result can be seen in Fig.~\ref{FigMag4}, in
which is plotted the reduced critical temperature versus the reduced
chemical potential. The full, dashed and dot-dashed lines represent
the critical lines for three values of the reduced applied field,
$\delta=0.01$, $\delta=0.5$ and $\delta=1.5$, respectively, at fixed
value of the reduced size, $\chi =0.5$. It suggests that the
critical temperature decreases as the chemical potential increases
and, as in Fig.~\ref{FigMag3}, the broken phase is strengthened for
stronger values of the applied field.

Finally, in Fig.~\ref{FigMag5} it is plotted again the reduced
critical temperature versus the reduced chemical potential, but for
three values of the reduced inverse size, $\chi=1.0$, $\chi =2.0$
and $\chi =2.5$, at the fixed value of the reduced applied field,
$\delta =0.1$ (respectively full, dashed and dot-dashed lines). We
see that the broken phase is inhibited for smaller sizes.

In summary, we have investigated how a magnetic background affects
the size-dependent phase structure of the scalar field theory in the
framework of 2PI formalism, in the Hartree--Fock approximation,
considering the large-$N$ limit. We have found that the broken phase
is strengthened for stronger values of the applied field. Also, the
minimal size of the system, below which there is no phase
transition, is smaller for greater values of the applied field. We
thus conclude that the magnetic field drives the system to the
broken phase.

\section*{ACKNOWLEDGMENTS}

The authors thank the Brazilian agencies CAPES, CNPq and FAPERJ, for
financial support.

\end{document}